\documentclass[12pt,a4paper,final]{iopart}
\usepackage{iopams}

\expandafter\let\csname equation*\endcsname\relax
\expandafter\let\csname endequation*\endcsname\relax

\usepackage{amsmath}
\usepackage{amsfonts}
\usepackage{amsmath}
\usepackage{amssymb}
\usepackage{amsthm}
\usepackage[utf8]{inputenc}
\usepackage{bm}
\usepackage{graphicx}
\usepackage{cite}

\usepackage[breaklinks=true,colorlinks=true,linkcolor=blue,urlcolor=blue,citecolor=blue]{hyperref}

\newcommand{\defeq}{\mathrel{\mathop:}=}

\newcommand{\U}{\mathcal{U}}
\renewcommand{\S}{\mathcal{S}}
\renewcommand{\H}{\mathcal{H}}

\newcommand{\D}{\mathcal{D}}
\newcommand{\vth}{v_{\text{th}}}

\begin{document}

\title{A superstatistical measure of distance from canonical equilibrium}

\author[cor1]{Sergio Davis$^{1,2}$}
\address{$^1$Research Center on the Intersection in Plasma Physics, Matter and Complexity (P$^2$mc), Comisión Chilena de Energía Nuclear, Casilla 188-D, Santiago, Chile}
\address{$^2$Departamento de F\'isica, Facultad de Ciencias Exactas, Universidad Andres Bello. Sazi\'e 2212, piso 7, 8370136, Santiago, Chile.}
\ead{sergio.davis@cchen.cl}

\begin{abstract}
Non-equilibrium systems in steady states are commonly described by generalized statistical mechanical theories such as non-extensive statistics and superstatistics. 
Superstatistics assumes that the inverse temperature $\beta = 1/(k_B T)$ follows some pre-established statistical distribution, however, it has been previously 
proved (Physica A \textbf{505}, 864-870 [2018]) that $\beta$ cannot be associated to an observable function $B(\bm \Gamma)$ of the microstates $\bm \Gamma$. In this 
work, we provide an information-theoretical interpretation of this theorem by introducing a new quantity $\D$, the mutual information between $\beta$ and 
$\bm \Gamma$. Our results show that $\D$ is also a measure of departure from canonical equilibrium, and reveal a minimum, non-zero uncertainty about $\beta$ given 
$\bm \Gamma$ for every non-canonical superstatistical ensemble. This supports the use of the mutual information as a descriptor of complexity and correlation in 
complex systems, also providing in some cases a sound basis for the use of Tsallis' entropic index $q$ as a measure of distance from equilibrium, being in those 
cases a proxy for $\D$.
\end{abstract}

\section{Introduction}

Complex systems, such as plasmas~\cite{Ourabah2015, Ourabah2020b, Davis2019b} and self-gravitating systems~\cite{Lima2002, Du2004b, Iguchi2005, Ourabah2020c}, as 
well as systems in economic~\cite{Tsallis2003, Denys2016} and social sciences~\cite{Prenga2021}, commonly deviate from the behavior predicted by the canonical ensemble, 
and in those cases are sometimes described using generalized statistical mechanical theories such as Tsallis' non-extensive statistics~\cite{Tsallis2009c}, superstatistics~\cite{Beck2003, Beck2004} among others. In particular, we are thinking of classical systems having a Hamiltonian $\H(\bm \Gamma)$ where 
$\bm \Gamma$ is a microstate, that is, a point in phase space. For instance, take a system composed of $N$ interacting particles such that 
$\bm \Gamma = (\bm{r}_1, \ldots, \bm{r}_N, \bm{p}_1, \ldots, \bm{p}_N)$, being $\bm{r}_i$ their positions and $\bm{p}_i$ their momenta, with a Hamiltonian of 
the form
\begin{equation}
\H(\bm \Gamma) = \sum_{i=1}^N \frac{\bm{p}_i^2}{2 m_i} + \Phi(\bm{r}_1, \ldots, \bm{r}_N),
\end{equation}
where $\Phi$ is an interaction energy, possibly long-range. 

If such a system is capable of reaching thermal equilibrium, then its microstates with be distributed according to the canonical ensemble,
\begin{equation}
\label{eq:canonical}
P(\bm \Gamma|\beta) = \frac{\exp\big(-\beta\H(\bm \Gamma)\big)}{Z(\beta)},
\end{equation}
with $\beta \defeq 1/(k_B T)$ the inverse temperature. However, in the case of complex systems this is not generally the case, and generalized ensembles have been 
proposed to describe them. A common generalization of \eqref{eq:canonical} is the $q$-canonical ensemble, usually associated with Tsallis statistics, and whose distribution 
of microstates is given by
\begin{equation}
\label{eq:qcanonical}
P(\bm \Gamma|q, \beta_0) = \frac{1}{Z_q(\beta_0)}\Big[1+(q-1)\beta_0 \H(\bm \Gamma)\Big]^\frac{1}{1-q},
\end{equation}
with $q$ the \emph{entropic index}. As is well known, the limit $q \rightarrow 1$ recovers the canonical ensemble. One particular case of $q$-canonical ensemble is the 
so-called kappa distribution~\cite{Livadiotis2017} of particle velocities in plasmas,
\begin{equation}
\label{eq:kappa1}
P(\bm v|\kappa, \vth) = \frac{1}{\eta_1}\left(1+\frac{1}{\kappa-\frac{3}{2}}\frac{v^2}{\vth^2}\right)^{-(\kappa+1)},
\end{equation}
where $\vth$ is a scale parameter for $\bm v$ known as the \emph{thermal velocity}, and the shape parameter $\kappa$ is referred to as the \emph{spectral index}. 
This index is related to the entropic index $q$ by the equivalence
\begin{equation}
q = 1 + \frac{1}{\kappa+1}.
\end{equation}

In non-equilibrium systems governed by these distributions, it is common practice to use the entropic index $q$ or equivalently, the spectral index $\kappa$ as a 
measure of complexity. However, a proper foundation of this identification on the principles of information theory remains to be developed. Since the last decade, 
the use of mutual information to quantify complexity and the presence of emergent properties has been considered by several groups both in classical and quantum 
contexts~\cite{Ball2010, Galla2012, Valdez2017, Varley2022, Navarrete2022}.

More recently, an impossibility theorem~\cite{Davis2018} was proved, revealing that the superstatistical inverse temperature $\beta$ cannot be associated to a phase 
space function $B(\bm \Gamma)$ in the same way that energy $E$ is associated to the value of the Hamiltonian function $\H(\bm \Gamma)$. Motivated by the need for a 
deeper understanding of the meaning of this theorem, in this work we introduce a new descriptor $\D$, corresponding to the mutual information between $\beta$ and 
the microstate variables $\bm \Gamma$, as a measure of distance from the canonical equilibrium within the superstatistical family of ensembles. We explore the 
behavior of this quantity by considering a model system of classical particles with velocities following the kappa distribution in \eqref{eq:kappa1}, and show 
that $\D$ is a strictly increasing function of the relative variance of $\beta$.

\newpage
\section{Superstatistics and the fundamental temperature}

Superstatistics~\cite{Beck2003, Beck2004} is the unique statistical mechanical framework in which the inverse temperature $\beta$ is promoted, from a fixed parameter, 
to a random variable. If $S$ denotes the set of parameters that define a particular superstatistical ensemble, then the probability density of $\beta$ for that ensemble 
will be written as $P(\beta|S)$. In this theory, \eqref{eq:canonical} is therefore replaced by the joint distribution
\begin{equation}
\label{eq:joint}
P(\bm \Gamma, \beta|S) = P(\bm \Gamma|\beta, S)P(\beta|S) = P(\bm \Gamma|\beta)P(\beta|S),
\end{equation}
where the first equality uses the product rule of probability theory, and the last equality assumes that knowledge of $S$ is irrelevant when $\beta$ is given, 
hence $P(\bm \Gamma|\beta, S)$ is replaced with $P(\bm \Gamma|\beta)$, which is the canonical ensemble in \eqref{eq:canonical}. In this way, the joint distribution 
in \eqref{eq:joint} reduces to
\begin{equation}
\label{eq:superstat}
P(\bm{\Gamma}, \beta|S) = \frac{\exp(-\beta \H(\bm \Gamma))}{Z(\beta)}P(\beta|S).
\end{equation}

The distribution of microstates $\bm \Gamma$ is given by integrating out $\beta$ according to the marginalization rule~\cite{Sivia2006} of probability theory,
\begin{equation}
\label{eq:superstat_marg}
P(\bm{\Gamma}|S) = \int_0^\infty d\beta P(\bm{\Gamma}, \beta|S) = \int_0^\infty d\beta P(\beta|S)\left[\frac{\exp\big(-\beta \H(\bm \Gamma)\big)}{Z(\beta)}\right],
\end{equation}
where we can see that $P(\bm \Gamma|S)$ depends on $\bm \Gamma$ only through the Hamiltonian $\H$. That is, there exists an \emph{ensemble function} $\rho(E; S)$ of 
the energy $E$ such that
\begin{equation}
\label{eq:rho}
P(\bm{\Gamma}|S) = \rho(\H(\bm \Gamma); S).
\end{equation}

\noindent
This ensemble function can be written as
\begin{equation}
\label{eq:rho_laplace}
\rho(E; S) = \int_0^\infty d\beta f(\beta; S)\exp(-\beta E),
\end{equation}
i.e. as the Laplace transform of the \emph{superstatistical weight function}
\begin{equation}
\label{eq:fbeta}
f(\beta; S) \defeq \frac{P(\beta|S)}{Z(\beta)}.
\end{equation}

In this way, superstatistics can be formulated in terms of the functions $\rho(E; S)$ and $f(\beta; S)$ for simplicity, always keeping in mind the underlying 
structure in terms of the actual probability densities $P(\beta|S)$ and $P(\bm \Gamma|S)$.

It is important to note, for the discussion in the following sections, that if the ensemble is canonical, then $\beta$ and $\bm \Gamma$ are statistically 
independent variables, as can be seen from \eqref{eq:superstat}. If we use the label $\beta_0$ instead of $S$ to denote a canonical ensemble at inverse 
temperature $\beta_0$, we have
\begin{equation}
P(\beta|\beta_0) = \delta(\beta-\beta_0),
\end{equation}
therefore
\begin{equation}
\begin{split}
P(\bm \Gamma, \beta|\beta_0) & = \frac{\exp(-\beta \H(\bm \Gamma))}{Z(\beta)}\delta(\beta-\beta_0) 
= \frac{\exp(-\beta_0 \H(\bm \Gamma))}{Z(\beta_0)}\delta(\beta-\beta_0) \\
& = P(\bm \Gamma|\beta_0)P(\beta|\beta_0),
\end{split}
\end{equation}
where $P(\bm \Gamma|\beta_0)$ and $P(\beta|\beta_0)$ are the marginal distributions. For any other superstatistical ensemble $S$, the variables $\bm \Gamma$ and 
$\beta$ are correlated, which means $\bm \Gamma$ carries information about $\beta$ and viceversa. This correlation can be assessed using the conditional distribution 
of $\beta$ given $\bm \Gamma$, obtained from \eqref{eq:superstat} and Bayes' theorem as
\begin{equation}
\label{eq:pbeta_gamma}
P(\beta|\bm{\Gamma}, S) = \frac{P(\beta|S)P(\bm \Gamma|\beta)}{P(\bm \Gamma|S)} = \frac{f(\beta; S)\exp(-\beta\H(\bm \Gamma))}{\rho(\H(\bm \Gamma); S)}.
\end{equation}

\noindent 
As shown in \ref{app:cond_mean}, the mean of this distribution is given by
\begin{equation}
\label{eq:cond1}
\big<\beta\big>_{\bm{\Gamma}, S} = \beta_F(\H(\bm \Gamma); S),
\end{equation}
where
\begin{equation}
\label{eq:betafund}
\beta_F(E; S) \defeq -\frac{\partial}{\partial E}\ln \rho(E; S)
\end{equation}
is the \emph{fundamental inverse temperature} function~\cite{Davis2023b}, while its variance is
\begin{equation}
\label{eq:cond2}
\big<(\delta \beta)^2\big>_{\bm{\Gamma}, S} = -{\beta_F}'(\H(\bm \Gamma); S).
\end{equation}

\noindent
A direct consequence of \eqref{eq:cond2} is that, for all energies $E$, the inequality
\begin{equation}
\label{eq:betafprime_ineq}
{\beta_F}'(E; S) \leq 0
\end{equation}
must hold for superstatistics. Taking expectation of \eqref{eq:cond1} under $S$ we see that
\begin{equation}
\begin{split}
\big<\beta_F\big>_S & = \int d\bm{\Gamma} P(\bm{\Gamma}|S)\big<\beta\big>_{\bm \Gamma, S} 
= \int_0^\infty d\beta \left[\int d\bm{\Gamma}P(\bm \Gamma|S)P(\beta|\bm \Gamma, S)\right]\beta \\
& = \int_0^\infty d\beta P(\beta|S)\beta = \big<\beta\big>_S,
\end{split}
\end{equation}
thus it makes sense to define the inverse temperature $\beta_S$ of the ensemble as
\begin{equation}
\beta_S \defeq \big<\beta_F\big>_S = \big<\beta\big>_S.
\end{equation}

\section{Temperature is not an observable in superstatistics}

In this section we will revisit a recently presented impossibility theorem~\cite{Davis2018}, which denies the existence of a universal, observable function 
$B(\bm \Gamma)$ that is interchangeable with the variable $\beta$ for any superstatistical ensemble $S$ other than the canonical. In this sense, temperature 
is not seen on the same footing as energy, where an observable function $\H(\bm \Gamma)$ does exist.

In more precise terms, what the existence of this observable inverse temperature $B(\bm \Gamma)$ would mean is that, for any function $G(\beta, \bm \Gamma)$ one 
should have the identity
\begin{equation}
\label{eq:theorem_cond}
\big<G(\beta, \bm \Gamma)\big>_S = \big<G(B(\bm \Gamma), \bm \Gamma)\big>_S.
\end{equation}

We will prove the impossibility of \eqref{eq:theorem_cond} by contradiction. Suppose that, for a given superstatistical ensemble $S$, a function $B(\bm \Gamma)$ 
exists such that \eqref{eq:theorem_cond} holds. Using the choice
\begin{equation}
G(\beta, \bm \Gamma) = \delta(\beta - \beta_0)\delta(\bm{\Gamma}-\bm{\Gamma}_0)
\end{equation}
in \eqref{eq:theorem_cond}, with $\bm{\Gamma}_0$ an arbitrary point in phase space and $\beta_0$ an arbitrary value of inverse temperature, we have
\begin{equation}
\label{eq:delta_delta}
\big<\delta(\beta-\beta_0)\delta(\bm{\Gamma}-\bm{\Gamma}_0)\big>_S = \big<\delta(B(\bm \Gamma)-\beta_0)\delta(\bm{\Gamma}-\bm{\Gamma}_0)\big>_S.
\end{equation}

\noindent 
Because of the ``expectation projection'' identity (see \ref{app:projection}),
\begin{equation}
\big<A\,\delta(\bm{\Gamma}-\bm{\Gamma}_0)\big>_S = P(\bm{\Gamma}_0|S)\cdot\big<A\big>_{\bm{\Gamma}_0, S}.
\end{equation}
for any quantity $A$, it follows from \eqref{eq:delta_delta} that
\begin{equation}
\begin{split}
P(\bm{\Gamma}_0|S)\cdot P(\beta_0|\bm{\Gamma}_0, S) & = P(\bm{\Gamma}_0|S) \cdot P(B = \beta_0|\bm{\Gamma}_0, S) \\
& = P(\bm{\Gamma}_0|S)\cdot\delta\big(\beta_0 - B(\bm{\Gamma}_0)\big),
\end{split}
\end{equation}
hence, by cancelling the factor $P(\bm{\Gamma}_0|S)$ from both sides, we have
\begin{equation}
\label{eq:pbeta_delta}
P(\beta|\bm{\Gamma}, S) = \delta\big(\beta-B(\bm \Gamma)\big),
\end{equation}
for any value of $\beta$ and $\bm \Gamma$. The result in \eqref{eq:pbeta_delta} is easily interpreted by the following argument. If $\beta$ is interchangeable with 
$B(\bm \Gamma)$, then perfect knowledge of $\bm \Gamma$ must guarantee perfect knowledge of $\beta$, which in turn must be represented by a Dirac delta distribution. 
Two conditions can be deduced from \eqref{eq:pbeta_delta}, namely that, for every $\bm \Gamma$, the expected value of $\beta$ given $\bm \Gamma$ should be 
$B(\bm \Gamma)$, that is,
\begin{equation}
\label{eq:Bmean}
\big<\beta\big>_{\bm{\Gamma}, S} = B(\bm \Gamma),
\end{equation}
and the variance of $\beta$ given $\bm \Gamma$ should be zero, i.e.
\begin{equation}
\label{eq:Bvar}
\big<(\delta \beta)^2\big>_{\bm{\Gamma}, S} = 0.
\end{equation}

\noindent
On the one hand, consistency of \eqref{eq:Bmean} with \eqref{eq:cond1} requires that 
\begin{equation}
B(\bm \Gamma) = \beta_F(\H(\bm \Gamma); S)
\end{equation}
for every $\bm \Gamma$, therefore, if $B(\bm \Gamma)$ does exist, it must be equal to the fundamental inverse temperature. On the other hand, consistency of 
\eqref{eq:Bvar} with \eqref{eq:cond2} requires that ${\beta_F}'(\H(\bm \Gamma); S) = 0$ for every $\bm \Gamma$, so the only possibility left is that the fundamental 
inverse temperature function is the constant function, $\beta_F(E; S) = \beta_0$ for all $E$, that is, $S$ is the canonical ensemble at inverse temperature $\beta_0$. 
We have then proved that the existence of $B(\bm \Gamma)$ interchangeable with $\beta$ is only allowed in the canonical limit of superstatistics.

An alternative sketch of a proof can be readily obtained using the \emph{inverse temperature covariance} $\U$, a quantity recently proposed~\cite{Davis2022b}. This 
quantity can be defined for any ensemble of the form in \eqref{eq:rho} by
\begin{equation}
\label{eq:U}
\U \defeq \big<(\delta \beta_F)^2\big>_S - \big<{\beta_F}'\big>_S,
\end{equation}
however, for superstatistics $\U$ simply becomes the variance of $\beta$. That is, in superstatistics it holds true that
\begin{equation}
\label{eq:U_ident}
\big<(\delta \beta_F)^2\big>_S - \big<{\beta_F}'\big>_S = \big<(\delta \beta)^2\big>_S.
\end{equation}

If we now replace $\beta$ by $B(\bm \Gamma)$ in \eqref{eq:cond1}, we again conclude that $\beta$ has to be interchangeable with $\beta_F$, and then \eqref{eq:U_ident} 
implies that
\begin{equation}
\big<{\beta_F}'\big>_S = 0.
\end{equation}

Now, because of \eqref{eq:betafprime_ineq}, the only possibility is that ${\beta_F}'(E; S) = 0$ for all energies $E$, thus $S$ can only be a canonical ensemble.

In other words, the theorem in Ref.~\cite{Davis2018} tells us the following: whenever there is uncertainty over the value of $\beta$ in the state $S$, even 
perfect knowledge of $\bm \Gamma$ cannot completely remove that uncertainty. We can confirm this by computing the relative variance of $\beta$ given $\bm \Gamma$, 
which must have a non-zero minimum value. For instance, the $q$-canonical ensemble in \eqref{eq:qcanonical} is described by superstatistics for $q \geq 1$, with 
$q = 1$ corresponding to the canonical ensemble. In this case we have
\begin{equation}
\beta_F(E; q, \beta_0) = -\frac{\partial}{\partial E}\ln \rho(E; q, \beta_0) = \frac{\beta_0}{1 + (q-1)\beta_0 E},
\end{equation}
therefore, using \eqref{eq:cond1} and \eqref{eq:cond2}, we see that
\begin{equation}
\frac{\big<(\delta \beta)^2\big>_{\bm \Gamma, q, \beta_0}}{\big<\beta\big>_{\bm \Gamma, q, \beta_0}^2} 
= -\frac{{\beta_F}'(\H(\bm \Gamma); q, \beta_0)}{\beta_F(\H(\bm \Gamma); q, \beta_0)^2} = q-1,
\end{equation}
which is larger than zero unless we are in the canonical ensemble ($q$ = 1).

As the uncertainty in $\beta$ does not disappear even when fully knowing $\bm \Gamma$, it must have a different origin. One possibility is that of a Bayesian 
interpretation of superstatistics, such as the one discussed in Refs.~\cite{Sattin2006, Sattin2018}, where the uncertainty in $\beta$ is not due to it being a 
fluctuating quantity, but rather an unknown one. On the other hand, it is possible to understand $\beta$ as a fluctuating quantity depending on degrees of freedom 
\emph{external} to $\bm \Gamma$, but correlated with it. In order to illustrate this point of view, let us take a composite system divided into two subsystems, 
a \emph{target} with degrees of freedom $\bm{\Gamma}_A$ and an \emph{environment} with degrees of freedom $\bm{\Gamma}_B$. The composite system has a joint 
distribution $P(\bm{\Gamma}_A, \bm{\Gamma}_B|S)$, and the marginal ensemble describing the target is
\begin{equation}
\label{eq:marg_A}
P(\bm{\Gamma}_A|S) = \int d\bm{\Gamma}_B P(\bm{\Gamma}_A|\bm{\Gamma}_B, S)P(\bm{\Gamma}_B|S).
\end{equation}

As shown in \ref{app:composite}, the target distribution $P(\bm{\Gamma}_A|S)$ can be expressed as a superstatistical ensemble if the conditional distribution $P(\bm{\Gamma}_A|\bm{\Gamma}_B, S)$ is of the form
\begin{equation}
\label{eq:ansatz}
P(\bm{\Gamma}_A|\bm{\Gamma}_B, S) = \left[\frac{\exp\big(-\beta \H_A(\bm{\Gamma}_A)\big)}{Z_A(\beta)}\right]_{\beta = B(\bm{\Gamma}_B)},
\end{equation}
which is the case discussed in Ref.~\cite{Davis2020}, where the probability density of the superstatistical variable $\beta$ is given by
\begin{equation}
\label{eq:pbeta_env}
P(\beta|S) = \Big<\delta\big(\beta-B(\bm{\Gamma}_B)\big)\Big>_S = \int d\bm{\Gamma}_B P(\bm{\Gamma}_B|S)\delta\big(\beta-B(\bm{\Gamma}_B)\big).
\end{equation}

Here $\beta$ can in fact be interchanged with an observable function $B$, but said function can only depend on the degrees of freedom of the environment. 

\section{A superstatistical distance from canonical equilibrium}

We have seen in the previous section that knowledge of $\bm \Gamma$ can never lead to perfect knowledge of $\beta$ in superstatistics, unless we are in the 
canonical ensemble. Nevertheless, knowing the microstate $\bm \Gamma$ should reduce the uncertainty on $\beta$, and this idea can be formalized using concepts 
from information theory, in particular entropy and mutual information.

There are several possibilities when attempting to define an entropy for a superstatistical system, as we must be careful of precisely agreeing on the set of 
degrees of freedom we are describing. Previous and recent works discussing the concept of entropy in the context of superstatistics give further insight on this 
issue~\cite{Abe2007, Beck2011, Ourabah2024}.

\noindent
Perhaps the first entropy one would define for $S$ is the entropy of the microstates,
\begin{equation}
\S_{\bm{\Gamma}}(S) \defeq -\int d\bm{\Gamma}P(\bm \Gamma|S)\ln P(\bm \Gamma|S) = -\big<\ln \rho(\H)\big>_S.
\end{equation}

This is the standard Boltzmann-Gibbs entropy in thermodynamics, only this time applied to the generalized ensemble $S$. The limit when $S$ becomes a canonical 
ensemble at $\beta_0$ is of course well defined, and is given by
\begin{equation}
\S_{\bm{\Gamma}}(\beta_0) = \beta_0\big<\H\big>_{\beta_0} + \ln Z(\beta_0).
\end{equation}

\noindent
Another relevant entropy may be the entropy of the variable $\beta$ itself, given by
\begin{equation}
\S_\beta(S) \defeq \big<-\ln P(\beta|S)\big>_S = -\int_0^\infty d\beta P(\beta|S)\ln P(\beta|S),
\end{equation}
although its use may be problematic, as $\S_\beta \rightarrow -\infty$ when approaching the canonical ensemble. On the other hand, the uncertainty about $\beta$ 
when knowledge of $\bm \Gamma$ is acquired is best represented by the conditional entropy $\S_{\beta|\bm{\Gamma}}$ of the distribution in \eqref{eq:pbeta_gamma},
\begin{equation}
\S_{\beta|\bm{\Gamma}}(\bm \Gamma, S) \defeq \big<-\ln P(\beta|\bm \Gamma, S)\big>_{\bm{\Gamma}, S} = -\int_0^\infty d\beta P(\beta|\bm \Gamma, S)\ln P(\beta|\bm \Gamma, S).
\end{equation}

We are now ready to approach the problem of defining a superstatistical distance from thermodynamic equilibrium, and for this we introduce the relative entropy 
of the microstates $\bm \Gamma$, from the canonical ensemble at inverse temperature $\beta_0$ to the actual superstatistical ensemble $S$. We define this relative 
entropy by
\begin{equation}
\S_{\bm{\Gamma}}(\beta_0 \rightarrow S) \defeq \left<-\ln \left[\frac{P(\bm \Gamma|S)}{P(\bm \Gamma|\beta_0)}\right]\right>_S
= -\int d\bm{\Gamma} P(\bm \Gamma|S)\ln \left[\frac{P(\bm \Gamma|S)}{P(\bm \Gamma|\beta_0)}\right].
\end{equation}

Note that $\S_{\bm \Gamma}(\beta_0 \rightarrow S)$ is defined with respect to a reference canonical ensemble at a given temperature, however, our aim is to define 
a descriptor such that it takes into account all possible values of $\beta$ in the superstatistical ensemble $S$. Therefore we instead introduce the quantity
\begin{equation}
\D(S) \defeq \left<-\ln \left[\frac{P(\bm \Gamma|S)}{P(\bm \Gamma|\beta)}\right]\right>_S
= -\int_0^\infty d\beta \int d\bm{\Gamma} P(\bm \Gamma, \beta|S)\ln \left[\frac{P(\bm \Gamma|S)}{P(\bm \Gamma|\beta)}\right].
\end{equation}

\noindent
Replacing $P(\bm \Gamma|S)$ according to \eqref{eq:rho} and $P(\bm \Gamma|\beta)$ from \eqref{eq:canonical}, we obtain
\begin{equation}
\D = \left<\ln \left[\frac{\exp(-\beta \H)}{Z(\beta)\rho(\H; S)}\right]\right>_S
= \left<\ln \left[\frac{P(\beta|S)\exp(-\beta \H)}{Z(\beta)P(\beta|S)\rho(\H; S)}\right]\right>_S,
\end{equation}
and by identifying $P(\bm \Gamma, \beta|S)$ according to \eqref{eq:superstat}, we can finally express $\D$ as
\begin{equation}
\D = \left<\ln \left[\frac{P(\bm \Gamma, \beta|S)}{P(\beta|S)P(\bm \Gamma|S)}\right]\right>_S.
\end{equation}

We recognize the right-hand side as the \emph{mutual information} between the variables $\beta$ and $\bm \Gamma$. The mutual information~\cite{CoverThomas2006} 
$I_{XY}$ is a measure of the amount of information that a random variable $X$ contains about another random variable $Y$, and it is defined, for a model with 
parameters $\boldsymbol{\theta}$, by
\begin{equation}
I_{XY}(\boldsymbol{\theta}) \defeq \left<\ln \left[\frac{P(X, Y|\boldsymbol{\theta})}{P(X|\boldsymbol{\theta})P(Y|\boldsymbol{\theta})}\right]\right>_{\boldsymbol{\theta}}.
\end{equation}

This can be interpreted as measuring the distance between the actual distribution and a model that assumes statistical independence between the variables. From 
Jensen's inequality, it can be shown that 
\begin{equation}
\label{eq:mutual_positive}
I_{XY}(\boldsymbol{\theta}) \geq 0
\end{equation}
with $I_{XY}(\boldsymbol{\theta}) = 0$ if and only if the variables are statistically independent, i.e., if and only if $P(X, Y|\boldsymbol{\theta}) 
= P(X|\boldsymbol{\theta})P(Y|\boldsymbol{\theta})$. Therefore, we can write
\begin{equation}
\label{eq:mutual_super}
\D \defeq I_{\beta\bm \Gamma} = \left<\ln \left[\frac{P(\bm{\Gamma}, \beta|S)}{P(\bm{\Gamma}|S)P(\beta|S)}\right]\right>_S \geq 0.
\end{equation}

We see that $\D$ is an alternative measure of distance from equilibrium as others previously proposed~\cite{LopezRuiz1995, Livadiotis2010, Pennini2017b}. After using 
the product rule in the form
\begin{equation}
P(\bm \Gamma, \beta|S) = P(\beta|\bm \Gamma, S)P(\bm \Gamma|S),
\end{equation}
we can alternatively express $\D$ as
\begin{equation}
\D = \left<\ln \left[\frac{P(\beta|\bm{\Gamma}, S)}{P(\beta|S)}\right]\right>_S = \S_\beta - \big<\S_{\beta|\bm{\Gamma}}\big>_S,
\end{equation}
hence, as $\D \geq 0$ we immediately have
\begin{equation}
\label{eq:sbeta_ineq}
\S_\beta \geq \big<\S_{\beta|\bm{\Gamma}}\big>_S,
\end{equation}
with equality ($\D$ = 0) only when $P(\beta|\bm \Gamma, S) = P(\beta|S)$, that is, according to \eqref{eq:pbeta_gamma}, when $P(\bm \Gamma|S) = P(\bm \Gamma|\beta)$, i.e. 
$S$ is a canonical ensemble. It is in the sense of the inequality in \eqref{eq:sbeta_ineq} that we can say that the uncertainty about $\beta$ is reduced when 
knowledge of $\bm \Gamma$ is acquired. 

In summary, if $S$ is a canonical state at an arbitrary inverse temperature $\beta_0$, then $\D = 0$, while for any other superstatistical state, 
$\D > 0$ and we say that the microstate $\bm{\Gamma}$ carries information about $\beta$ (and viceversa). However, this information will never be enough to 
exactly determine the value of $\beta$, according to the theorem previously proved.

\section{An example: particles with kappa-distributed velocities}

As an illustration, let us consider a system composed of $N$ particles having a joint distribution of velocities given by
\begin{equation}
\label{eq:kappan}
P(\bm{v}_1, \ldots, \bm{v}_N|u, \beta_S) = \frac{1}{\eta_N}\left(1+u\beta_S\sum_{i=1}^N \frac{m_i \bm{v}_i^2}{2}\right)^{-\big(\frac{1}{u}+\frac{3N}{2}\big)}
\end{equation}
with $0 \leq u \leq \frac{1}{2}$ and where $\eta_N$ is a normalization constant given by
\begin{equation}
\eta_N = \left(\frac{2\pi}{u\beta_S}\right)^{\frac{3N}{2}}\frac{\Gamma\Big(\frac{1}{u}\Big)}{\Gamma\Big(\frac{3N}{2}+\frac{1}{u}\Big)}
\prod_{i=1}^N m_i^{-\frac{3}{2}}.
\end{equation}

This follows the treatment of the kappa distribution used in Ref.~\cite{Davis2023e}, which was obtained from superstatistics as the ensemble function 
\begin{equation}
\label{eq:kappan_rho}
\rho_N(K; u, \beta_S) = \int_0^\infty d\beta \left[\frac{P(\beta|u, \beta_S)}{Z_N(\beta)}\right]\exp(-\beta K)
= \frac{1}{\eta_N}\Big(1+u\beta_S K\Big)^{-\big(\frac{1}{u}+\frac{3N}{2}\big)},
\end{equation}
with
\begin{equation}
Z_N(\beta) = \left(\frac{2\pi}{\beta}\right)^{\frac{3N}{2}}\;\prod_{i=1}^N m_i^{-\frac{3}{2}}
\end{equation}
the partition function of an $N$-particle ideal gas. The probability density $P(\beta|u, \beta_S)$ is a gamma distribution,
\begin{equation}
\label{eq:pbeta}
P(\beta|u, \beta_S) = \frac{1}{u\beta_S\Gamma(1/u)}\exp\left(-\frac{\beta}{u\beta_S}\right)\left(\frac{\beta}{u\beta_S}\right)^{\frac{1}{u}-1}.
\end{equation}

From the first and second moment of \eqref{eq:pbeta} we see that $\beta_S = \big<\beta\big>_S$ and $u$ is the relative variance of $\beta$,
\begin{equation}
u = \frac{\big<(\delta \beta)^2\big>_{u,\beta_S}}{\big<\beta\big>_{u, \beta_S}^2}.
\end{equation}

\noindent
The value of $u$ is connected to the spectral index $\kappa$ via
\begin{equation}
\kappa = \frac{1}{u} + \frac{1}{2},
\end{equation}
so that $u \rightarrow 0$ (equivalent to $\kappa \rightarrow \infty$) reproduces the Maxwellian distribution for individual particle velocities. The entropy of 
the distribution of $\beta$ in \eqref{eq:pbeta} is
\begin{equation}
\label{eq:entropy_beta}
\S_\beta = \frac{1}{u} + \psi(1/u) - \phi(1/u) + \ln \Gamma(1/u) + \ln\;(u\beta_S),
\end{equation}
where $\psi(z)$ is the digamma function, and we have defined for convenience the function
\begin{equation}
\phi(z) \defeq z\psi(z).
\end{equation}

\noindent
Knowledge of the kinetic energy $K$ yields the conditional distribution
\begin{equation}
\label{eq:pbeta_givenK}
P(\beta|K, u, \beta_S) = \frac{\big[1+u\beta_S K\big]^{\frac{1}{u}+\frac{3N}{2}}}{ u\beta_S\Gamma(\frac{1}{u}+\frac{3N}{2}) }\exp\left(-\frac{\beta}{u\beta_S}\big[1+u\beta_S K\big]\right)\left(\frac{\beta}{u\beta_S}\right)^{\frac{1}{u}+\frac{3N}{2}-1},
\end{equation}
which is also a gamma distribution, having a relative variance given by
\begin{equation}
\frac{\big<(\delta \beta)^2\big>_{K, u, \beta_S}}{\big<\beta\big>_{K, u, \beta_S}^2} = \frac{2u}{2+3Nu} \leq u.
\end{equation}

Here we see that adding knowledge of $K$ always decreases the relative variance of $\beta$ unless $u = 0$, which is the only case in which this relative variance 
can be zero for finite $N$. The conditional entropy $\S_{\beta|\bm{V}}$ only depends on $\bm V=(\bm{v}_1, \ldots, \bm{v}_N)$ through the kinetic energy $K(\bm V)$, 
so we have
\begin{equation}
\S_{\beta|\bm{V}}(\bm V, u, \beta_S) = \zeta(K(\bm V); u, \beta_S),
\end{equation}
with
\begin{equation}
\label{eq:zeta}
\zeta(K; u, \beta_S) \defeq -\int_0^\infty d\beta P(\beta|K, u, \beta_S)\ln P(\beta|K, u, \beta_S).
\end{equation}

\noindent
Replacing \eqref{eq:pbeta_givenK} and after some calculation, we obtain
\begin{equation}
\label{eq:Scond_betaV}
\zeta(K) = \ln\Big(\frac{u\beta_S}{1+u\beta_S K}\Big) - \phi\Big(\frac{3N}{2}+\frac{1}{u}\Big) + \psi\Big(\frac{3N}{2}+\frac{1}{u}\Big)
+ \frac{3N}{2}+\frac{1}{u}+\ln\Gamma\Big(\frac{3N}{2}+\frac{1}{u}\Big).
\end{equation}

\begin{figure}[t!]
\begin{center}
\includegraphics[width=0.7\textwidth]{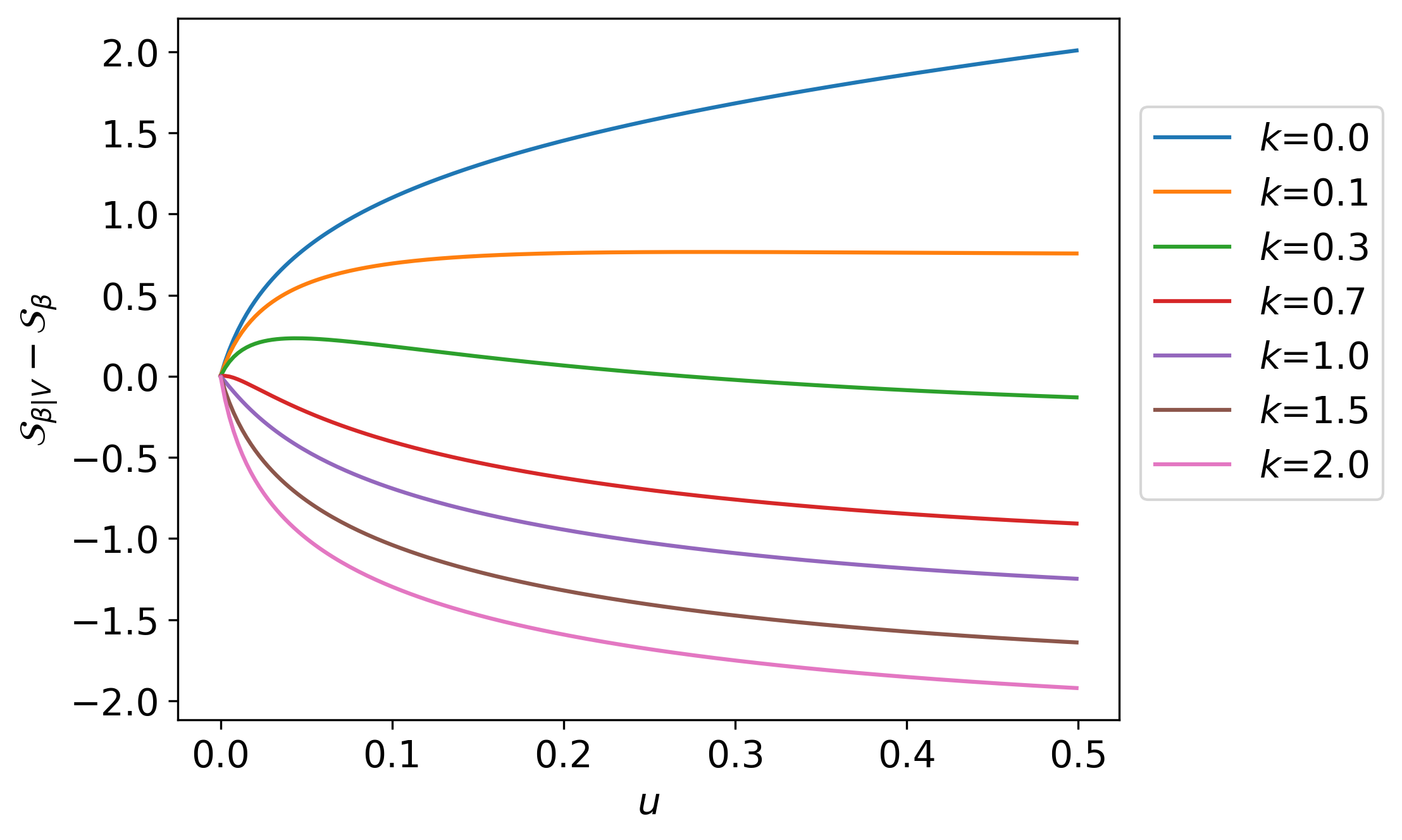}
\end{center}
\caption{Difference between the conditional entropy $\S_{\beta|\bm{V}}=\zeta(K(\bm V))$ with $\zeta$ in \eqref{eq:zeta} and the entropy $\S_\beta$ in 
\eqref{eq:entropy_beta} as a function of $u$ for $\beta_S = 1$, $N$ = 50 and different values of $k = K/N$.}
\label{fig:entropydiff}
\end{figure}

Fig.~\ref{fig:entropydiff} shows the difference between $\S_{\beta|\bm V}$ in \eqref{eq:Scond_betaV} and the entropy $\S_\beta$ in \eqref{eq:entropy_beta}. We can 
see that the difference vanishes in the limit $u \rightarrow 0$, and in fact we can verify that
\begin{equation}
\lim_{u \rightarrow 0} \S_{\beta|\bm V} = \lim_{u \rightarrow 0} \Big[\ln\;(u\beta_S) - \phi(1/u) + \frac{1}{u} + \ln \Gamma(1/u)\Big] 
= \lim_{u \rightarrow 0} \S_\beta.
\end{equation}

\noindent
The mutual information $\D(u, \beta_S)$ between $\beta$ and $\bm{V}$ is then
\begin{equation}
\D(u, \beta_S) = \S_\beta(u, \beta_S) - \big<\S_{\beta|\bm{V}}\big>_{u,\beta_S},
\end{equation}
and upon replacing in the second term on the right-hand side
\begin{equation}
\Big<\ln\; (1+u\beta_S K)\Big>_{u, \beta_S} = \psi\Big(\frac{3N}{2}+\frac{1}{u}\Big) - \psi\Big(\frac{1}{u}\Big),
\end{equation}
obtained by computing the expected value of $\ln (1+u\beta_S K)$ under the distribution
\begin{equation}
\begin{split}
P(K|u, \beta_S) & = \int_0^\infty d\beta P(\beta|u, \beta_S)\left[\frac{\beta^{\frac{3N}{2}}\exp(-\beta K)K^{\frac{3N}{2}-1}}{\Gamma(\frac{3N}{2})}\right] \\
& = \frac{u\beta_S\Gamma(\frac{3N}{2}+\frac{1}{u})}{\Gamma(\frac{3N}{2})\Gamma(\frac{1}{u})}\big[1+u\beta_S K\big]^{-\left(\frac{1}{u}+\frac{3N}{2}\right)} (u\beta_S K)^{\frac{3N}{2}-1},
\end{split}
\end{equation}
we finally arrive at
\begin{equation}
\label{eq:mutual_kappa}
\D(u; N) = \ln \Gamma\left(\frac{1}{u}\right)-\ln \Gamma\left(\frac{3N}{2}+\frac{1}{u}\right) + \phi\left(\frac{3N}{2}+\frac{1}{u}\right)-\phi\left(\frac{1}{u}\right)
-\frac{3N}{2}.
\end{equation}

\begin{figure}[h!]
\begin{center}
\includegraphics[width=0.495\textwidth]{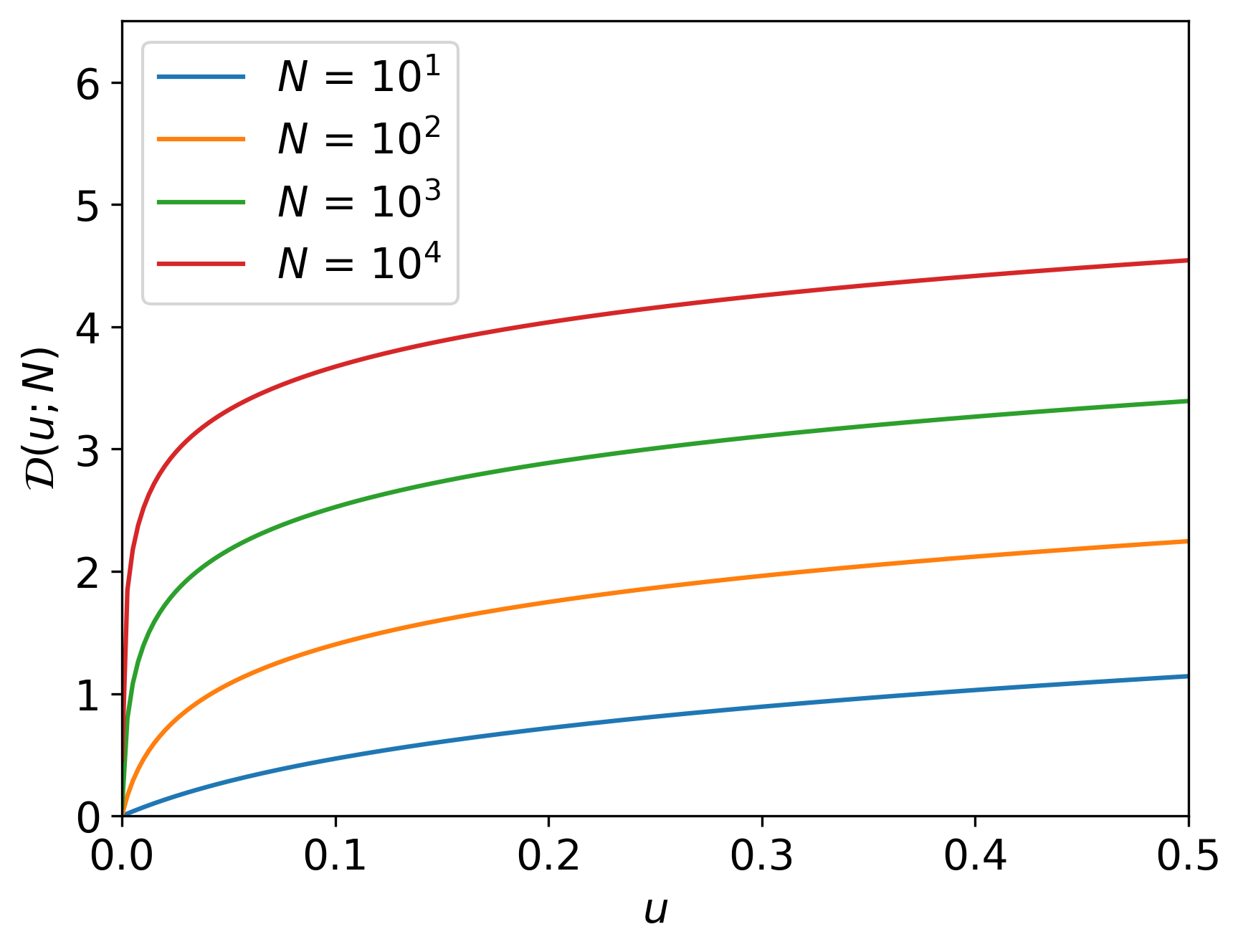}
\includegraphics[width=0.495\textwidth]{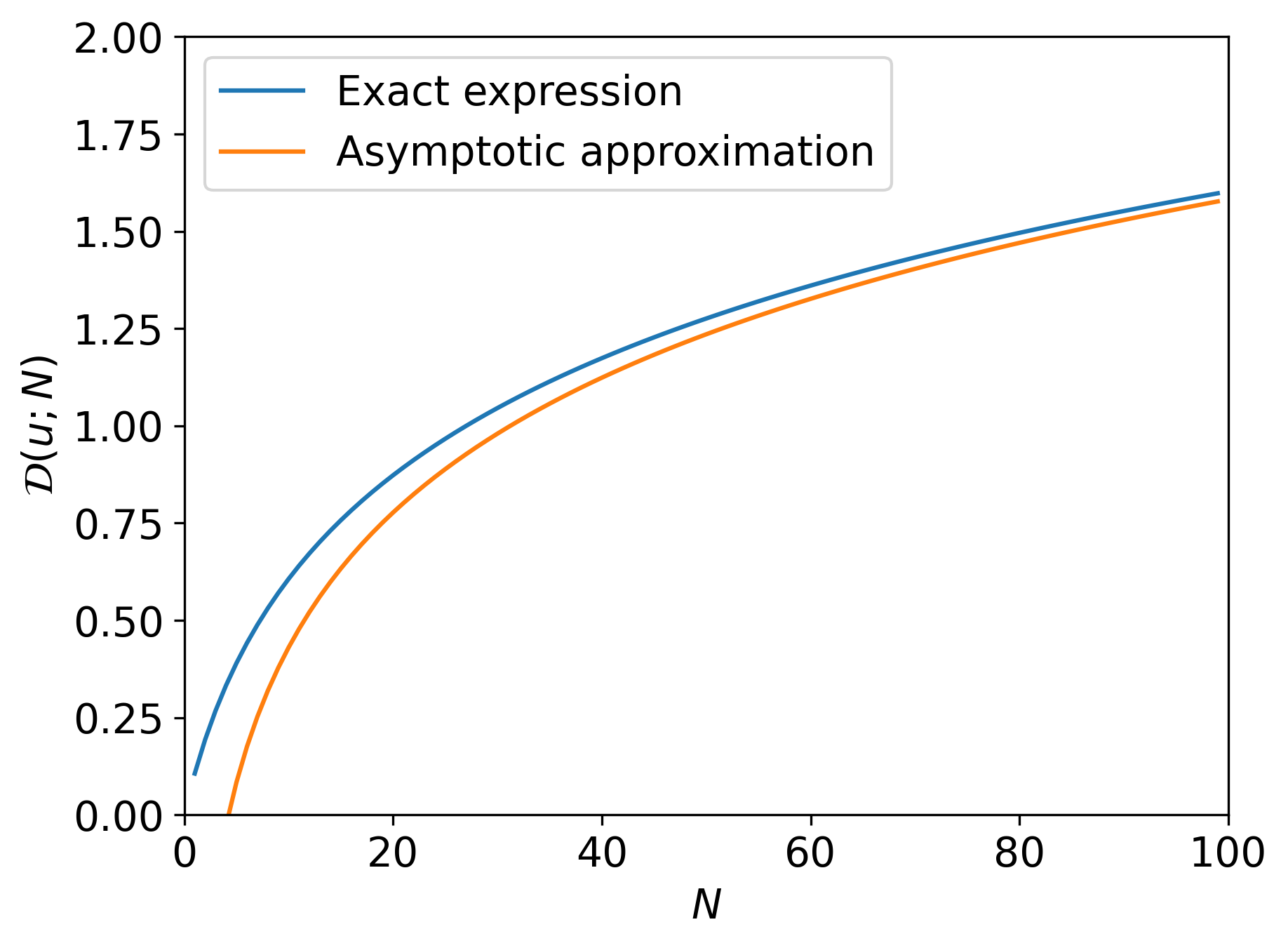}
\end{center}
\caption{Mutual information function $\D(u; N)$ as given by \eqref{eq:mutual_kappa}. Left panel, $\D(u; N)$ as a function of $u$ for different values of $N$. Right 
panel, $\D(u; N)$ as a function of $N$ for $u$ = 0.15 versus the asymptotic approximation in \eqref{eq:asymptotic}.}
\label{fig:mutual}
\end{figure}

It is interesting to note that, unlike $\S_\beta$ and $\S_{\beta|\bm{V}}$, $\D$ does not depend on the mean inverse temperature $\beta_S$, being in fact a 
non-negative, monotonically increasing function of $u$, as seen in Fig.~\ref{fig:mutual} (left panel), with $\D$ = 0 only for $u$ = 0 as expected. Its derivative 
is given by
\begin{equation}
\frac{\partial \D(u; N)}{\partial u} = \frac{1}{u^2}\left[\frac{1}{u}\psi'\Big(\frac{1}{u}\Big) - \Big(\frac{3N}{2}+\frac{1}{u}\Big)\psi'\Big(\frac{3N}{2}+\frac{1}{u}\Big)\right],
\end{equation}
where $\psi'(z) \defeq d\psi(z)/dz$ is the trigamma function, and we see that $\partial D/\partial u$ is non-negative, as the function $z\mapsto z\psi'(z)$ 
is monotonically decreasing.

\noindent
Furthermore, $\D(u; N)$ is a strictly increasing function of $N$ for fixed $u$, with derivative
\begin{equation}
\frac{\partial \D(u; N)}{\partial N} = \frac{3}{2}\left[\Big(\frac{3N}{2}+\frac{1}{u}\Big)\psi'\Big(\frac{3N}{2}+\frac{1}{u}\Big)-1\right] > 0,
\end{equation}
as can be seen in Fig.~\ref{fig:mutual} (right panel). For large $N$, $\D(u; N)$ has the asymptotic approximation
\begin{equation}
\label{eq:asymptotic}
\D(u; N) \sim \frac{1}{2}\ln\;(3N) + \frac{1}{u} + \ln \Gamma(1/u) - \phi(1/u) - \ln 2 - \frac{1}{2}(1 + \ln \pi),
\end{equation}
showing that $\D$ is non-extensive, as its growth for $N \rightarrow \infty$ is slower than $N$.

\noindent
For a single particle following a kappa distribution, we can write $\D$ in terms of $\kappa$ as
\begin{equation}
\label{eq:DS_kappa_1}
\D(\kappa) = \ln \Gamma\Big(\kappa-\frac{1}{2}\Big)-\ln \Gamma(\kappa+1) + \phi(\kappa+1)-\phi\Big(\kappa-\frac{1}{2}\Big)-\frac{3}{2},
\end{equation}
for which it holds that $\D(\kappa) \rightarrow 0$ as $\kappa \rightarrow \infty$.

\section{Concluding remarks}

We have presented in a new light the impossibility theorem of Ref.~\cite{Davis2018}, which denies the existence of a function $B(\bm \Gamma)$ interchangeable with 
$\beta$ in superstatistics, by uncovering its significance in terms of information theory. In brief, the theorem is equivalent to the statement that there is a 
minimum, non-zero uncertainty on $\beta$ given $\bm \Gamma$ in all superstatistical ensembles except for the canonical. Furthermore, knowledge of $\bm \Gamma$ 
only translates into information about $\beta$ (assuming the ensemble $S$ is known) if $S$ is not canonical, information that can be quantified using the mutual 
information $I_{\beta\bm \Gamma}(S)$ between $\beta$ and $\bm \Gamma$, a non-negative quantity which and coincides with the measure $\D$ of distance from canonical 
equilibrium to the ensemble $S$. In short, the further away a superstatistical system is from equilibrium, the more information $\bm \Gamma$ carries about the temperature.

\newpage
We have also shown, in the case of classical particles with kappa-distributed velocities, that the use of $q$ and $\kappa$ as measures of departure from equilibrium can 
be justified by the fact that $\D(u, \beta_S)$ does not depend on $\beta_S$ and is a strictly increasing function of $u$, with $\D$ = 0 only for $u = 0$.

\section*{Acknowledgments}

SD gratefully acknowledges funding from ANID FONDECYT 1220651 grant.

\appendix

\section{Conditional mean and variance of the inverse temperature for a given microstate}
\label{app:cond_mean}

Taking the logarithmic derivative of \eqref{eq:rho_laplace} with respect to $E$, we see that
\begin{equation}
\label{eq:dlogrho}
\frac{\partial}{\partial E} \ln \rho(E; S) = -\int_0^\infty d\beta \left[\frac{f(\beta; S)\exp(-\beta E)}{\rho(E; S)}\right]\beta
\end{equation}
and from the definition of $\beta_F$ in \eqref{eq:betafund},
\begin{equation}
\beta_F\big(\H(\bm \Gamma); S\big) = \left[-\frac{\partial}{\partial E}\ln \rho(E; S)\right]_{E = \H(\bm \Gamma)}
= \int_0^\infty d\beta \left[\frac{f(\beta; S)\exp\big(-\beta \H(\bm \Gamma)\big)}{\rho\big(\H(\bm \Gamma); S\big)}\right]\beta.
\end{equation}

\noindent
Replacing \eqref{eq:pbeta_gamma} we obtain
\begin{equation}
\label{eq:betaF_app}
\beta_F\big(\H(\bm \Gamma); S\big) = \int_0^\infty d\beta P(\beta|\bm \Gamma, S)\beta = \big<\beta\big>_{\bm \Gamma, S}.
\end{equation}

\noindent
On the other hand, differentiating \eqref{eq:dlogrho} with respect to $E$ we have
\begin{equation}
\frac{\partial^2}{\partial E^2}\ln \rho(E; S) = -\left(\frac{\partial}{\partial E}\ln \rho(E; S)\right)^2 + \int_0^\infty d\beta 
\left[\frac{f(\beta; S)\exp(-\beta E)}{\rho(E; S)}\right]\beta^2,
\end{equation}
therefore, replacing $E$ by $\H(\bm \Gamma)$ and using \eqref{eq:betaF_app}, we obtain
\begin{equation}
{\beta_F}'\big(\H(\bm \Gamma); S\big) = \big<\beta\big>_{\bm \Gamma, S}^2 -\big<\beta^2\big>_{\bm \Gamma, S} = -\big<(\delta \beta)^2\big>_{\bm \Gamma, S}.
\end{equation}

\section{Proof of the expectation projection identity}
\label{app:projection}

Let us explicitly write the desired expectation value $\big<A\;\delta(\bm \Gamma - \bm{\Gamma}_0)\big>_S$ as an integral in terms of the underlying random variables, 
which in this case are $A$ itself and $\bm \Gamma$. As the relevant distribution is $P(A, \bm \Gamma|S)$ we have
\begin{equation}
\begin{split}
\big<A\;\delta(\bm{\Gamma}-\bm{\Gamma}_0)\big>_S & = \int dA\cdot A\int d\bm{\Gamma} P(A, \bm{\Gamma}|S)\;\delta(\bm{\Gamma}-\bm{\Gamma}_0) \\
& = \int dA\cdot A\;P(A, \bm{\Gamma}_0|S).
\end{split}
\end{equation}

\noindent
Using the product rule as $P(A, \bm{\Gamma}_0|S) = P(A|\bm{\Gamma}_0, S)\;P(\bm{\Gamma}_0|S)$ we can further write
\begin{equation}
\big<A\;\delta(\bm{\Gamma}-\bm{\Gamma}_0)\big>_S = P(\bm{\Gamma}_0|S)\int dA\cdot A\;P(A|\bm{\Gamma}_0, S) = P(\bm{\Gamma}_0|S)\big<A\big>_{\bm{\Gamma}_0, S}.
\end{equation}

\section{Superstatistical treatment for a region of a composite system}
\label{app:composite}

In order to see how \eqref{eq:marg_A} becomes a superstatistical ensemble, we introduce into its right-hand side a factor of 1 as the integral
\begin{equation}
\int_0^\infty d\beta \delta\big(\beta-B(\bm{\Gamma}_B)\big) = 1,
\end{equation}
and replace \eqref{eq:ansatz}, then obtaining
\begin{equation}
\begin{split}
P(\bm{\Gamma}_A|S) & = \int d\bm{\Gamma}_B\left[\int_0^\infty d\beta \delta\big(\beta-B(\bm{\Gamma}_B)\big)\right] P(\bm{\Gamma}_A|\bm{\Gamma}_B, S)P(\bm{\Gamma}_B|S) \\
& = \int d\bm{\Gamma}_B \int_0^\infty d\beta\left[\frac{\exp\big(-B(\bm{\Gamma}_B)\H_A(\bm{\Gamma}_A)\big)}{Z_A(B(\bm{\Gamma}_B))}\right]
\delta\big(\beta-B(\bm{\Gamma}_B)\big) P(\bm{\Gamma}_B|S) \\
& = \int d\bm{\Gamma}_B \int_0^\infty d\beta\left[\frac{\exp\big(-\beta \H_A(\bm{\Gamma}_A)\big)}{Z_A(\beta)}\right]
\delta\big(\beta-B(\bm{\Gamma}_B)\big) P(\bm{\Gamma}_B|S),
\end{split}
\end{equation}
where in the last line we have replaced $B(\bm{\Gamma}_B)$ by $\beta$ in the expression inside the square brackets by virtue of the Dirac delta. Now, changing the order 
of integration and replacing $P(\beta|S)$ as per \eqref{eq:pbeta_env}, we finally obtain our desired result,
\begin{equation}
P(\bm{\Gamma}_A|S) = \int_0^\infty d\beta P(\beta|S)\left[\frac{\exp\big(-\beta \H_A(\bm{\Gamma}_A)\big)}{Z_A(\beta)}\right],
\end{equation}
which is a superstatistical ensemble according to \eqref{eq:superstat_marg}.

\section*{References}

\bibliography{mutualinf}
\bibliographystyle{unsrt}

\end{document}